

\magnification=\magstep1
\baselineskip=14pt
\parindent=20pt
\parskip=0pt

\pageno=1

\def\mydate{{May 25, 1995}}


\font\ninerm=cmr9

\font\bfbig=cmb10  scaled\magstep1

\font\nineit=cmti9

\font\eightsl=cmsl8

\font\ninebf=cmbx9

\def\mybig{\displaystyle \strut }

\def\d{\partial}
\def\la{\raise.16ex\hbox{$\langle$}\lower.16ex\hbox{}  }
\def\ra{\, \raise.16ex\hbox{$\rangle$}\lower.16ex\hbox{} }
\def\go{\rightarrow}

\def\onehalf{ \hbox{${1\over 2}$} }
\def\psibar{ \psi \kern-.65em\raise.6em\hbox{$-$} }
\def\mbar{ m \kern-.75em\raise.4em\hbox{$-$}\hbox{} }
\def\delbar{{\bar\delta}}

\def\mstar{$\lower.18em\hbox{*}$~}
\def\mystar{\lower.18em\hbox{*}}

\def\ep{\epsilon}

\def\vphi{ {\varphi} }

\def\mass{{\rm mass}}


\def\boxit#1{\vbox{\hrule\hbox{\vrule\kern3pt
     \vbox{\kern3pt\hbox{#1}\kern3pt}\kern3pt\vrule}\hrule}}
\def\bigbox#1{\vbox{\hrule\hbox{\vrule\kern5pt
     \vbox{\kern5pt\hbox{#1}\kern5pt}\kern5pt\vrule}\hrule}}
\def\Bigbox#1{\vbox{\hrule\hbox{\vrule\kern8pt
     \vbox{\kern8pt\hbox{#1}\kern8pt}\kern8pt\vrule}\hrule}}

\def\midbox#1{\hbox{\vbox{\hsize=5cm \hrule\hbox{\vrule\kern5pt
     \vbox{\kern5pt\hbox{#1}\kern5pt}\kern5pt\vrule}\hrule}}}


\line{\ninerm \mydate \hfil UMN-TH-1347/95}
\vglue .6cm

\centerline{\bfbig More About the Massive Multi-flavor Schwinger
Model\footnote{$^\dagger$}{\ninerm To appear in the Proceedings of {\nineit
the Nihon University Workshop on Fundamental Problems in Particle Physics},
March 31 - April 1, 1995}}

\vskip .5cm
\centerline{Yutaka Hosotani}
\vskip 5pt
{\baselineskip=10pt
\centerline{\eightsl School of Physics and Astronomy, University of Minnesota}
\centerline{\eightsl Minneapolis, Minnesota 55455, USA}
}
\vskip .3cm

{\ninerm  \baselineskip=11pt
\midinsert  \narrower\narrower \noindent
The massive multi-flavor Schwinger model on a circle is reduced to
a finite dimensional quantum mechanics problem.  The model sensitively
depends on the parameter $mL|\cos\onehalf\theta|$ where $m$, $L$, and $\theta$
are a typical fermion mass, the volume, and the vacuum angle, respectively.
\endinsert
}

\vskip .3cm

The Schwinger model is QED in two dimensions.  Its Lagrangian is given by
$$ {\cal L} = - \hbox{$1\over 4$} \, F_{\mu\nu} F^{\mu\nu} +
\sum_{a=1}^N \psibar_a \Big\{ \gamma^\mu (i \d_\mu - e A_\mu) -
  m_a  \Big\}    \psi_a  ~. \eqno(1)  $$
Schwinger showed [1] in 1962 that for $N=1$ and $m=0$, the gauge boson acquires
a mass $\mu$ given by  $\mu^2 = e^2 / \pi$.  The original massless Schwinger
model is exactly solvable.  It admits the chiral condensate $\la
\psibar\psi \ra \not= 0$, the $\theta$ vacuum, and the confinement
phenomenon.  With a non-vanishing fermion mass  the model is not exactly
solvable.  Still, if $m \ll \mu$, the model can be approximately solved.

When the number of flavor $N$ is more than one, something puzzling happens.
Coleman analyzed the $N=2$ massive Schwinger model in 1976.[2]  With the aid
of the bosonization technique, he showed that there appear two bosons.  The
first one is essentially the same as the gauge boson in the $N=1$ model.  For
$m_1=m_2\ll e$, its mass squared is doubled: $\mu_1^2 = 2 e^2/\pi$.  A
surprising finding was the second boson picks up a mass given by
$$
\mu_2 \sim  m^{2/3} \mu_1^{1/3} \big| \cos \onehalf \theta \big|^{2/3}
\hskip 1cm {\rm (Coleman ~76)}.  \eqno(2) $$
The expression involves fractional powers of  $m$ and $| \cos \onehalf
\theta|$.  It is  essentially ``nonperturbative'' in nature.   This prompts
us to ponder. Why can't we apply a perturbation theory in fermion masses?

If all fermion masses vanish, the model (1) is exactly solvable for arbitrary
$N$.  The $U(1)$ chiral symmetry is broken by anomaly.
It is known that there result one massive boson ($\mu_1^2=Ne^2/\pi$) and
$N-1$ massless bosons.  $SU(N)$ chiral symmetry remains unbroken, but
correlators show power-law decay behavior.

We analyze the model, placing it on a circle.  There are new features which
are absent on a line or in the Minkowski spacetime.  First the constant mode
of gauge field $A_1$ becomes a dynamical variable, which we call the Wilson
line phase $\theta_W$.   It has been known that $\theta_W$ couples
to fermion degrees of freedom through chiral anomaly, leading to the
$\theta$ vacuum.[3]  Secondly, we are going to show that constant modes (zero
modes) of fermions along the circle play a decisive role in determining the
structure of the vacuum when $N>1$.  It will be seen that the puzzle about
(2) is also solved naturally on the way.

The result presented in this report is based on the work done in
collaboration with Jim Hetrick and Satoshi Iso.[4]

\def\squ{\kern-.1em\lower.25em\hbox{*}}
\def\stress{~ {}\squ\squ\squ ~}

The model carries several parameters.   Let's take $m_a=m$ for the moment.
Then the parameters are $e$, $m$, $\theta$ (vacuum angle), and $L$.
In the Minkowski spacetime limit $L\go \infty$, there are only two
dimensionless parameters:  $\big( m/e , \theta \big)$.  On a circle there
appears one more dimensionless parameter, say, $mL$.  We shall see
that the theory sensitively depends on this new parameter.    More precisely
speaking,
$$
\vcenter{ \Bigbox{ $
{}~ \stress  mL |\cos \onehalf\theta | \gg 1  \stress
   ~~ \not= ~~
    \stress  mL |\cos \onehalf\theta |\ll 1  \stress ~
$ } }
      \eqno(3)   $$

\noindent
This is a highly nontrivial statement.  It implies that limits $L\go\infty$,
$m\go 0$, and $\theta \go \pm \pi$ do not commute with each other.
If one takes the Minkowski limit $L\go\infty$ with $m\not=0$ and
$|\theta|<\pi$, then one is always in the regime on the left side.  However,
if we take a large, but finite $L$ and take the massless fermion limit
$m\go 0$, then one will be in the regime on the right side.  The massless
limit in the Minkowski spacetime is singular, although it is smooth so long
as $L$ is kept finite.

To see these, we first adopt the bosonization on a circle.  Suppose that
fermion fields $\psi_a=(\psi^a_+,\psi^a_-)$ obey boundary conditions
$\psi_a(t,x+L) = -\, e^{2\pi i \alpha_a } \, \psi_a(t,x)$.  We bosonize in
the interaction picture:
$$\eqalign{
&\psi^a_\pm(t,x) = { 1\over\sqrt{L}}
\, C^a_\pm \,
 e^{\pm i \{ q^a_\pm + 2\pi p^a_\pm (t \pm x)/L \} }
  :\, e^{\pm i\phi^a_\pm (t,x) } \, :  \cr
&C^a_+ =\exp \Big\{ i\pi \sum_{b=1}^{a-1} ( p^b_+ + p^b_- - 2 \alpha_b)
   \Big\}  ~~,~~
C^a_- = \exp \Big\{ i\pi \sum_{b=1}^{a} ( p^b_+ - p^b_-) \Big\} ~. \cr}
   \eqno(4)  $$
Here bosonic variables satisfy
$$\eqalign{
&[q^a_\pm, p^b_\pm] = i \, \delta^{ab}  ~~~,~~~
[c^a_{\pm,n}, c^{b,\dagger}_{\pm,m}] = \delta^{ab} \delta_{nm} \cr
&\phi^a_\pm (t,x) = \sum_{n=1}^\infty {1\over \sqrt{n}} \,
  \Big\{ c^a_{\pm,n} \, e^{- 2\pi in(t \pm x)/L} + {\rm h.c.} \Big\}~.\cr}
    \eqno(5)  $$
The boundary conditions are implemented by imposing the physical state
condition
$$e^{2\pi i p^a_\pm} ~ | \, {\rm phys} \ra = e^{2\pi i \alpha_a} ~
|\, {\rm phys} \ra  ~~.  \eqno(6)   $$

The only physical gauge field degree of freedom is the Wilson line phase
$e^{i\theta_W} = e^{ie \int_0^L dx \, A_1}$.   The Hamiltonian written in the
Schr\"odinger picture becomes
$$\eqalign{
H ~&= ~ H_0 ~+ ~H_\chi~ + ~H_{\rm mass} \cr
H_0 &= -{\pi \mu^2 L\over 4} {d^2\over d\theta_W^2}
+ {\pi\over 2L} \sum_{a=1}^N \Big\{ (p^a_+-p^a_-)^2
+ (p^a_+ + p^a_- + {\theta_W\over \pi} )^2 \Big\} \cr
H_\chi &= \int_0^L dx \, {1\over 2}
\bigg\{  N_\mu[ {\Pi_1}^2 + \chi_1'^2   + \mu^2 \,  \chi_1^2]
+ \sum_{\alpha=2}^N N_0[\Pi_\alpha^2 + \chi_\alpha'^2 ] \bigg\}   \cr}
   \eqno(7)   $$
Here $\chi_1= N^{-1/2} \sum_a (\phi^a_++\phi^a_-)$ represents the $U(1)$
`charge' part of oscillatory modes.  $\chi_2$ $\sim$ $\chi_N$ are other
orthogonal combinations.  Because of the Coulomb interaction $\chi_1$
acquires a mass-squared $\mu^2 = Ne^2/\pi$.   $H_{\rm mass}$ represents
the fermion mass term.

It immediately follows from the bosonized form (7) that the massless fermion
theory is trivial.  It is a free theory.  $H_0$ contains a non-trivial
coupling resulting from  the chiral anomaly, but is bilinear in dynamical
variables so that it can be solved exactly.   In particular, the oscillatory
mode part $H_\chi$ consists of 1 massive boson and $N-1$ massless bosons.
These $N-1$ massless bosons generate $SU(N)$ current algebra.

The presence of the mass term makes the theory nontrivial, and indeed
very rich.  Let's restrict ourselves to the $N=2$ (two flavor) case.
Eigenstates of $H_0$ are
$$\eqalign{
\Phi_s^{(n,r)} &= {1\over (2\pi)^2} \,
u_s(\theta_W + 2\pi n + \pi r + \pi \alpha_1 +\pi\alpha_2)
e^{i(n+\alpha_1)(q^1_++q^1_-) + i(n+r+\alpha_2)(q^2_++q^2_-)} \cr
E_s^{(n,r)}
&=  ~ \mu s + {\pi \over L} (r-\alpha_1+\alpha_2)^2 +  {\rm const.}   \cr}
   \eqno(8)  $$
Here $u_s(x)$ is the $s$-th eigenfunction in the harmonic oscillator problem.
The energy does not depend on $n$ due to the invariance under large gauge
transformations.  Consequently it is more natural to take $\theta$
eigenstates $\Phi^r_s(\theta) =  \sum_n e^{in\theta} \, \Phi_s^{(n,r)}$.

If the mass term $H_{\rm mass}$ is absent and $\alpha_1=\alpha_2$, then
$\Phi^0_0(\theta)$ is the lowest energy state, namely the vacuum.   It follows
that  $\la \psibar_a \psi_a \ra_\theta = 0$.

Now let us look at effects of the mass term  $H_{\rm mass} =
\int_0^L dx  \sum_a m_a \psibar_a\psi_a $.  Note
$\psi^{a\dagger}_- \psi^a_+ \propto  e^{i ( q^a_- + q^a_+ )}
{}~ N_0[ e^{i\sqrt{2\pi} \chi_1} ]   \, N_0[ e^{\pm i\sqrt{2\pi} \chi_2} ] $.
Hence,  $H_\mass$ gives  (1) transitions among various $\Phi^r_s(\theta)$'s,
(2) a mass for $\chi_2$, and (3) other interactions.  Notice that we need to
shift the reference mass of the normal product from
$ N_0[ e^{i\alpha \chi_a(x)} ]$ to
$N_{\mu_a}[ e^{i\alpha \chi_a(x)} ]$ where
$\mu_a$ is the physical mass of $\chi_a$ to be determined.

Having in mind that we are mostly interested in physics in a large
volume, we, for the moment, suppress transitions to higher $s$ states.
Then the mass term gives transitions among  $\Phi^{(n,r)}$'s:
$$\eqalign{
\Phi^{(n,r)}
{}~ &\longrightarrow\kern-1.7em\raise.6em\hbox{$m_1$}\lower.6em\hbox{}
{}~~~ \Phi^{(n\pm 1,r\mp 1)} \cr
&\longrightarrow\kern-1.7em\raise.6em\hbox{$m_2$}\lower.6em\hbox{}
{}~~~\Phi^{(n,r\pm 1)}  \cr}  $$
 Our strategy is to determine all matrix elements of
$H_\mass$ in the basis of $\Phi^{(n,r)}$, and diagonalize $H_0+H_\mass$
exactly.

Indeed, this can be done without approximation.
A close examination shows that the two mass parameters $m_1$ and $m_2$
enter in the combination of
$$\eqalign{
&m_1e^{-i\theta} + m_2 ~=~
\mbar (\theta) ~ e^{+ i \delbar(\theta)}  \cr
{\rm for~~} m_1=m_2, \quad
&\mbar = 2m \, \big| \cos \onehalf \theta \big|  ~~~,~~~
\delbar = -\onehalf \theta
  + \pi \Big[ {\theta+\pi \over 2\pi}  \Big]  \cr} \eqno(9)  $$
$\delbar(\theta)$ has a discontinuity at $\theta=\pm\pi$ where $\mbar(\theta)$
vanishes.

Exact eigenstates of the total Hamiltonian are sought in the form
$$\eqalign{
&| \Phi(\theta) \ra
= \int_0^{2\pi} d\vphi
    ~ f(\vphi + \delbar) \, | \Phi(\theta;\vphi) \ra \cr
& \Phi(\theta,\phi)  = {1\over 2\pi} \sum_{n,r} e^{in\theta+ ir\phi}
 \Phi^{(n,r)}  \cr}
   \eqno(10)  $$
The equation $(H_0+H_\mass) | \Phi(\theta) \ra = (\pi \ep /L) | \Phi(\theta)
\ra $ reads
$$\vcenter{
\Bigbox{~$\bigg\{  \Big(i {\mybig d\over \mybig d\vphi}
-\delta\alpha \Big)^2
   -  \kappa \, \cos \vphi \,\bigg\}
 f(\vphi) = \ep  \, f(\vphi)  $~}}   \eqno(11) $$
where
$$\eqalign{
&\delta\alpha=\alpha_1-\alpha_2  ~~,  \cr
\noalign{\kern 6pt}
&\kappa = {2\over \pi} \, \mbar L \, B(\mu_1 L)^{1/2}
  B(\mu_2 L)^{1/2} e^{-\pi/2\mu L} ~~, \cr
\noalign{\kern 5pt}
&\hskip 1cm B(\mu L) = \cases{1 &at ~$\mu L=0$\cr
    {\mybig \mu L e^\gamma \over\mybig 4\pi} &for $\mu L \gg 1$~. \cr}
\cr}
   \eqno(12) $$

Eq.\ (11) is  nothing but the equation describing a quantum pendulum !
The solution is controlled by the parameters
$\kappa$   and $\delta\alpha$.  In a large volume $\kappa \gg
1$ so that the  cosine potential term in (11) dominates.  $f(\vphi)$ is
localized around $\vphi=0$.
$$f(\vphi) \sim
 \exp \Big(-i\delta\alpha\vphi - \sqrt{{\displaystyle \kappa  \over
   \mybig  8}}~  \vphi^2 \Big)
\qquad {\rm for}~~ \kappa \gg 1 ~.   \eqno(13) $$
On the other hand, in a small volume limit $\kappa \ll 1$.
$f(\vphi)$ shows sensitive dependence on the boundary condition parameters:
$$f(\vphi) \sim \cases{
1 + {\mybig\kappa\over\mybig 1-4\delta\alpha^2}
(\cos \vphi - 2i\delta\alpha \sin\vphi)
       &for ${\mybig \kappa \over\mybig 1\pm2\delta\alpha} \ll 1$\cr
{\mybig 1\over \mybig\sqrt{2}} (1+ e^{\mp i\vphi})
+{\mybig\kappa\over\mybig 4\sqrt{2}} (e^{\pm i\vphi} + e^{\mp 2i\vphi} )
  &for $\delta\alpha=\pm \onehalf$, $\kappa \ll 1$~.\cr} \eqno(14) $$

The mass $\mu_2$ of the $\chi_2$ field is determined by
$$\vcenter{
\bigbox{~$\mu_2^2 = {\mybig 2\pi^2\over\mybig L^2} \, \kappa \,
 \mybig \int d\vphi \, \cos \vphi \, |f(\vphi)|^2 ~$}} \eqno(15)  $$
where the normalization $\int  d\vphi \, |f|^2=1$ is understood.
Notice that Eq.\ (15) must be solved self-consistently.  $f(\vphi)$ is
determined as
a function of $\kappa$, but $\kappa$ in (12) depends on $\mu_2$.  In some
limiting cases one can find analytic expressions.  Suppose that
$\delta\alpha=0$ and $m\ll \mu$ so that $\mu_1 \sim \mu$.  Then
$$\mu_2 = \cases{
4\sqrt{2}\,  m|\cos\onehalf\theta|\, e^{-\pi/2\mu L} &for  $\mu L \ll 1$~,\cr
\noalign{\kern 8pt}
4\sqrt{2} \,  m|\cos\onehalf\theta|\,\Big({\mybig \mu L e^\gamma\over\mybig
4\pi} \Big)^{1/2}
     &for $\mu L \gg 1 \gg m L (\mu L)^{1/2}|\cos\onehalf\theta|$~,\cr
\noalign{\kern 8pt}
(4e^{2\gamma} \,  m^2|\cos\onehalf\theta|^2 \mu )^{1/3}
    &for $mL (\mu L)^{1/2}|\cos\onehalf\theta| \gg 1$~.\cr}
   \eqno(16)  $$
This contains Coleman's result (2) in the case
$mL (\mu L)^{1/2}|\cos\onehalf\theta| \gg 1$.
Similarly the chiral condensate is found to be
$$\la \psibar\psi \ra_\theta = - {\mu_2^2\over 4\pi m} ~~.  \eqno(17) $$

We recognize that there is no puzzle about the fractional power dependence.  If
a mass $m$ is sufficiently small such that $m L (\mu
L)^{1/2}|\cos\onehalf\theta| \ll 1$,
then $\mu_2 \propto m$.  A perturbation theory may be employed.   An important
criterion is not whether $m/\mu$ is large or samll, but instead whether the
parameter $\kappa$ is large or small.  Depending on the value of $\kappa$, the
behavior of physical quantities is quite different.   This is the content of
the statement (3) above.

Furthermore, it is evident that all physical quantities are periodic in
$\theta$ with a period $2\pi$:

\centerline{\Bigbox{Periodicity in $\theta$ ~ = ~ $2\pi$}}

\noindent  This is so, even though $\onehalf\theta$-dependence is everywhere.
The appearance of the absolute value ensures the $2\pi$ periodicity.  There
appears no indication for the existence of fractional topological number
and `$4\pi$' periodicity which  have been claimed in the literature.[5]

Our analysis can be generalized.  One can take into account effects of
transitions to higher $s$ states which we have ignored above.  We end up
with an equation similar to Eq.\ (11), but this time with one more
variable describing dynamics in the $s$-direction.  The equation can be
solved numerically.

The $\delta\alpha$ dependence of various physical quantities can also be
evaluated numerically.  It is found that $|f(\vphi)|^2$
shows significant $\delta\alpha$ dependence for $\kappa <0.3$, whereas
the dependence becomes almost undetectable  for $\kappa > 2$.
In particular, in the massless fermion limit $mL\go 0$ but with a large $L$
$$\la \psibar\psi\ra_\theta = \cases{
   0 &for $\delta\alpha=0$\cr
   - \Big( {\mybig \mu e^\gamma\over \mybig 4\pi L}\Big)^{1/2}
   |\cos\onehalf\theta|  &for $\delta\alpha=\onehalf$\cr}   \eqno(18)  $$

\bigskip

In this survey I described a powerful method of analysing the Schwinger model.
Pictorially

\vskip 10pt
\centerline{\bigbox{Massive $N$-flavor  Schwinger Model}}

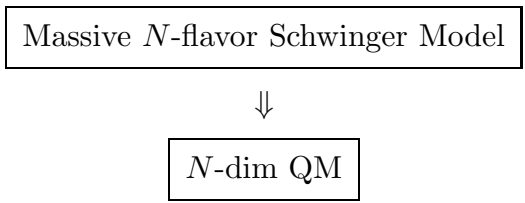
\centerline{$\Downarrow$}
\vskip 5pt

\centerline{\bigbox{$N$-dim QM}}
\vskip 5pt

\noindent
With this reduction one can solve the model to desired accuracy at least
numerically.

There are many intrigueing questions to be answered.
\itemitem{1.}  For $N=2$ we have observed that $\la \psibar\psi\ra_\theta$
vanishes at $\theta=\pi$.  Does it vanish at $\theta=2\pi/N$ in general?
\itemitem{2.}  Is the periodicity in $\theta$ always $2\pi$?
\itemitem{3.}  What happens in the chiral Schwinger model?  The above method
can be applied with a little modification.  Do we have fermion number
non-conservation?
\itemitem{4.}  What happens if each fermion has distinct charge such that the
ratio of two charges is irrational?  There is no $\theta$ vacuum in a rigorous
sense, but physics shouldn't depend on the charge parameters so sensitively.

\noindent
I will come back to these points shortly.

\vskip 1.cm

\def\ap {{\nineit Ann.\ Phys.\ (N.Y.)} }
\def\plB {{\nineit Phys.\ Lett.} {\ninebf B}}
\def\pr {{\nineit Phys.\ Rev.} }
\def\prD {{\nineit Phys.\ Rev.} {\ninebf D}}

{\parindent=15pt   \ninerm  \baselineskip=12pt
\leftline{\sl References:}
\vskip 3pt

\item{[1]}   J.\ Schwinger, \pr {\ninebf 125} (1962) 397;  {\ninebf 128} (1962)
2425.

\item{[2]}  S.\ Coleman,  \ap {\ninebf 101} (1976) 239.

\item{[3]}  J.E.\ Hetrick and Y.\ Hosotani, \prD {\ninebf 38} (1988) 2621.

\item{[4]}  J.E.\ Hetrick, Y.\ Hosotani, and S.\ Iso,
  hep-th/9502113, to appear in \plB .

\item{[5]} M.A.\ Shifman and A.V.\ Smilga, \prD {\ninebf 50} (1994) 7659.

}
\end

$$\eqalign{
(n, r&+1) \cr
&\kern-.2em\hbox{$\uparrow$}~ m_2\cr
(n-1,r+1)~ \longleftarrow ~ \kern-1.9em\raise.6em\hbox{$m_1$}\lower.6em\hbox{}
{}~~~(n~ ,& ~ r) ~~~

 ~~ (n+1, r-1) \cr
&\kern-.2em\hbox{$\downarrow$} ~ m_2\cr
(n, r & -1) \cr}
$$